\documentclass[%
 reprint,
superscriptaddress,
 amsmath,amssymb,
 aps,
]{revtex4-2}

\usepackage{graphicx}
\usepackage{dcolumn}
\usepackage{bm}

\usepackage[T1]{fontenc}

\usepackage{graphicx}
\usepackage{color}
\usepackage{amsmath}
\usepackage{upgreek}
\usepackage[left,modulo]{lineno}
\usepackage{xcolor}
\usepackage{ulem}

\begin{document}


\title{Slip and friction mechanisms at polymer semi-dilute solutions / solid interfaces}

\author{Marion Grzelka}
\affiliation{Universit\'e Paris-Saclay, CNRS, Laboratoire de Physique des Solides, 91405, Orsay, France}
\author{Iurii Antoniuk}
\affiliation{Univ Lyon, Universit\'e Lyon 1, CNRS, Ing\'enierie des Mat\'eriaux Polym\`eres, UMR 5223, F-69003, Lyon, France}
\author{Eric Drockenmuller}
\affiliation{Univ Lyon, Universit\'e Lyon 1, CNRS, Ing\'enierie des Mat\'eriaux Polym\`eres, UMR 5223, F-69003, Lyon, France}
\author{Alexis Chennevi\`ere}
\affiliation{Laboratoire L\'eon Brillouin, CEA Saclay, 91191 Gif-sur-Yvette, France}
\author{Liliane L\'eger}
\affiliation{Universit\'e Paris-Saclay, CNRS, Laboratoire de Physique des Solides, 91405, Orsay, France}
\author{Fr\'ed\'eric Restagno}
\email{frederic.restagno@universite-paris-saclay.fr}
\affiliation{Universit\'e Paris-Saclay, CNRS, Laboratoire de Physique des Solides, 91405, Orsay, France}
\altaffiliation{Laboratoire L\'eon Brillouin, CEA Saclay, 91191 Gif-sur-Yvette, France}


\date{\today}

\begin{abstract}
The role of the polymer volume fraction, $\phi$, on steady state slippage and interfacial friction is investigated for a semi-dilute polystyrene solutions in diethyl phthalate in contact with two solid surfaces. Significant slippage is evidenced for all samples, with slip lengths $b$ obeying a power law dependence. The Navier's interfacial friction coefficient, $k$, is deduced from the slip length measurements and from independent measurements of the solutions viscosity $\eta$. The observed scaling of $k$ versus $\phi$ clearly excludes a molecular mechanism of friction based on the existence of a depletion layer. Instead, we show that the data of $\eta(\phi)$ and $k(\phi)$ are understood when taking into account the dependence of the solvent friction on $\phi$. Two models, based on the friction of blobs or of monomers on the solid surface, well describe our data. Both points out that the Navier's interfacial friction is a semi-local phenomenon.

\end{abstract}

\maketitle

\section{Introduction}

Forcing a polymer fluid, melt or solutions, to flow past a solid wall is frequently encountered in a number of practical situations, ranging from polymer extrusion \cite{denn_extrusion_2001}, surface coating, or in microfluidic applications \cite{stone_microfluidics_2001}.  In order to optimize the corresponding practical processes, it is of out-most importance to be able to predict the exact flow behavior, in terms of velocity profile and of boundary condition for the flow velocity at the solid wall.  Contrary to simple fluids, where it is usually assumed that the fluid velocity at the fluid-solid interface is equal to zero or small for classical flow rates \cite{lauga_microfluidics_2005,neto_boundary_2005}, a finite and measurable flow velocity at the wall (or slip velocity) appears for polymer fluids \cite{migler_slip_1993,mhetar_slip_1998,wang_stick-slip_1996,baumchen_slip_2009,granick_slippery_2003,sanchez-reyes_interfacial_2003}. Indeed, when the fluid slides on the solid, the surface friction can be an important contribution to the global dissipation in the system.

\begin{figure}[htbp]
  \centering
    \includegraphics[width=0.6\columnwidth]{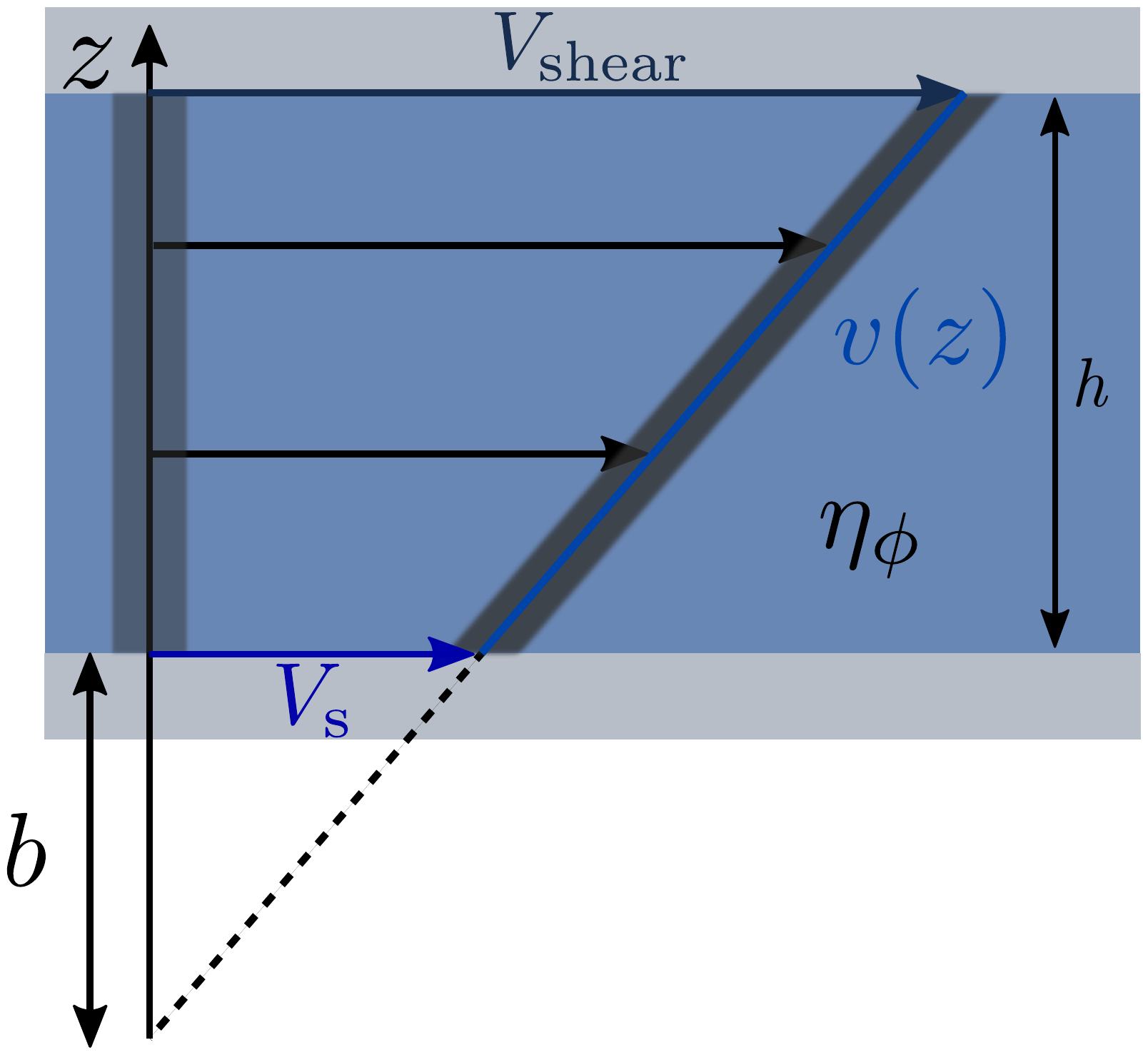}
    \caption{Definition of the slip length $b$ measured by velocimetry using photobleaching. A drop of homogeneous polymer solution of viscosity $\eta_\phi$ and thickness $h$, sheared at a velocity $V_\mathrm{shear}$, slips on the bottom surface with a slip velocity $V_\mathrm{s}$. The length at which the extrapolation of the velocity profile $v(z)$ reaches zero is called the slip length $b$.}
  \label{1}
\end{figure}

To characterize the friction at the solid interface, two classical ways are described in the literature. The most common one is to measure the slip length $b$, which is the distance in the solid where the velocity profile would extrapolate to zero (see Fig~\ref{1}). Another approach proposed by Navier \cite{navier_memoire_1823}, is to characterize the friction on the wall by a friction coefficient $k$ and assume that the sliding of the fluid on the surface at the the slip velocity $V_\text{s}$ as defined in Fig.~\ref{1}, creates a friction stress $\tau=k V_\text{s}$. The continuity of stresses at the solution/solid interface allows one to write the slip length as:
 \begin{equation}
    b=\frac{\eta}{k}~.
 \label{eq:b_definition}
 \end{equation}
 For polymer melts, the physical picture of the friction when the fluid is flowing on an ideal surface is now well established, and the interfacial friction appears to be a local phenomenon, involving length scales comparable to the size of a monomer. This idea, proposed by de Gennes in his seminal work \cite{de_gennes_ecoulements_1979}, gives a physical understanding of the large slip velocities reported in the literature for polymer melts \cite{reiter_real-time_2000,piau_measurement_1994,el_kissi_different_1990,mcgraw_slip-mediated_2016}. It allowed to predict a slip length of polymer melts proportional to $M_\mathrm{w}^3$, where $M_\mathrm{w}$ is the molar mass \cite{baumchen_slip_2009}. It also allowed one to understand the observed temperature dependencies of the slip length of polymer melts \cite{henot_comparison_2017} or the surprising equality between the friction coefficient measured for elastomers and melts made from the same chemical monomers \cite{henot_friction_2018}. The case of polymer melts flowing on surfaces with either grafted chains or adsorbed chains has also been extensively studied both experimentally \cite{durliat_influence_1997,ilton_adsorption-induced_2018,muller_flow_2008,montfort_polymer_2008} and theoretically \cite{gay_grafted_1996,brochard-wyart_slippage_1996,deng_simulation_2012} and the overall picture seems largely understood.

The situation appears more confused for polymer solutions. For dilute solutions, where the chains are not overlapping, the general picture for non-adsorbing surfaces is that a depletion layer develops close to the surface, so that pure solvent contacts the surface. The thickness of this depletion layer can be affected by the shear due to hydrodynamic interactions \cite{barnes_review_1995,graham_fluid_2011,ma_theory_2005} but the generally admitted idea is that no noticeable slip is expected for dilute solutions. Despite its strong interest in many situations such as enhanced oil recovery, cosmetics, food industry, the case of semi-dilute polymer solutions still remains poorly understood. Only few experimental data are available \cite{boukany_molecular_2010, archer_delayed_1995, archer_delayed_1995,sanchez-reyes_interfacial_2003,grzelka_viscoelasticity-induced_2020, barraud_boundary_2018,barraud_large_2019}, and they lead to contradictory conclusions.

We present below the results of an extensive investigation of the slip and friction mechanisms at semi-dilute good solvent polymer solutions/solid interfaces. We used a recently developed velocimetry technique, based on the monitoring of the flow-induced deformation of a thin line printed into the fluid by local fluorescence photobleaching \cite{henot_comparison_2017}, which is an optimization of that initially developed by Migler \textit{et al.} \cite{migler_slip_1993,durliat_influence_1997}. With the aim of gaining insight into the basic molecular mechanisms of interfacial friction, we have defined model systems composed of a series of solutions of a well-defined ultra-high molar mass polystyrene (PS, $M_\mathrm{n} = 10.2\,\mathrm{Mg}\cdot\mathrm{mol}^{-1}$, Ð =1.08), in a non-volatile good solvent (i.e. diethyl phthalate, DEP), and of two different model solid surfaces  (i.e. silicon substrates with and without a dense grafted layer of short PS chains).
The paper is organized as follows: after a brief description of the materials and methods, the results for the dependence of the slip length $b$, on the polymer volume fraction in the solution $\phi$, are presented for the two surfaces. These data, along with direct measurements of the solutions viscosities allow one to determine the polymer volume fraction dependence of the Navier's interfacial coefficient $k$. The observed scaling are compared to available theoretical models and provide precise lines of thought for a discussion on the molecular origin of the friction in these systems. In polymer semi-dilute solutions, mostly solvent is in contact with the solid wall. However, contrary to polymer melts were the interfacial friction is driven by local phenomenon involving monomer sizes, in solutions the strong dependence of the interfacial friction on the polymer volume fraction, appears to be compatible with the idea that it is mainly driven by the screening length of hydrodynamic interactions. This opens a new route to finely manipulate interfacial friction in such systems.

\section{Materials and Methods}

\textbf{Polymer solutions}

Series of semi-dilute solutions of a high molar mass polystyrene (PS) with chosen polymer volume fraction $\phi$ in diethyl phthalate (DEP), a non-volatile solvent, have been used, so that the polymer volume fraction could be kept constant during the duration of the flow characterization experiments. The protocol used to prepare the solutions is detailed in \cite{grzelka_viscoelasticity-induced_2020}. We briefly recall here its main steps. PS ($M_\mathrm{n} = 10.2\,\mathrm{Mg}\cdot\mathrm{mol}^{-1}$, \textit{\DJ}~= 1.08, Polymer Source Inc.) and ca. $1\mathrm{wt}\%$  of a photobleachable
polystyrene (PS di-NBD, $M_\mathrm{n} = 429 \,\mathrm{kg}\cdot\mathrm{mol}^{-1}$, \textit{\DJ} = 1.05) were dissolved in diethyl phthalate. Toluene was used as a co-solvent to accelerate the dissolution. Solutions were gently stirred for at least 3 weeks before evaporating the toluene under vacuum at room temperature for at least one week. The photobleachable labeled polystyrene contains nitro-benzoxadiazole (NBD) fluorescent groups emitting at $550\,\mathrm{nm}$ when excited at $458\,\mathrm{nm}$ at both chain ends. The synthesis and characterization of PS di-NBD are detailed in the SI of \cite{grzelka_viscoelasticity-induced_2020}. The Newtonian viscosity $\eta$ and the terminal relaxation (reptation) time $\tau_\mathrm{rep}$ of each solution were measured by oscillatory rheology at $22\,^\circ\mathrm{C}$ using an Anton-Paar MCR 302 rheometer in a cone plate geometry ($2\,^\circ$ cone angle, $25\,\mathrm{mm}$ diameter) (see Fig.~S3). Small angle neutron scattering experiments were performed in order to check 1) that DEP is indeed a good solvent for PS and 2) that the prepared solutions were in the semi-dilute concentration regime (see Supporting information). The blob size  could thus be determined in a large range of temperatures: $T\in[10-55]\,^\circ$C (see  Fig.~S2). Table \ref{tab6:solutions_PS} summarizes the characteristics of the PS in DEP solutions samples used in the present study.

 \begin{table}[htbp]
\centering
\begin{tabular}{c c c c}
\hline\hline
     $ \phi$ & $ \tau_\mathrm{rep}$ [s] & $ \eta$ [Pa$\cdot$s]  & $ \xi$ [\AA] \\
\hline
    0.0230 & 2.7 &  64 & 61 \\

    0.0314 & 8.3 & 401 & 48\\

    0.0397 & 16.3 & 1,147 & 40 \\

    0.0495 & 24.3 & 3,840 & 34 \\

    0.0608 & 50 & 12,000 & 29\\
\hline\hline
\end{tabular}
\caption{Characteristics of semi-dilute solutions of PS ($M_\mathrm{n}=10.2\,\mathrm{Mg}\cdot\mathrm{mol}^{-1}$, \textit{\DJ}=1.08) in DEP. The viscosity $\eta$ and reptation time $\tau_\mathrm{rep}$ have been measured by oscillatory rheology. The blob size $\xi$ is deduced from the SANS measurements, through the  scaling law $\xi=a\phi^{-0.75}$ with $a=0.36\,\text{nm}$ the effective size of a PS monomer unit.}
\label{tab6:solutions_PS}
\end{table}

\textbf{Solid substrates}

Two different model substrates were used: a bare silicon wafer and a silicon wafer covered with a dense layer of grafted-to short PS chains. The bare silicon wafer (2" diameter, $3\,\mathrm{mm}$ thick, Si-Mat Inc.) was cleaned by UV/O$_3$ treatment for at least $30\,\mathrm{min}$ before each measurement. The grafted PS layer was prepared following the protocol detailed in \cite{chenneviere_direct_2016}. The grafted PS ($M_\mathrm{n}=5.0\,\mathrm{kg}\cdot\mathrm{mol}^{-1}$, \textit{\DJ}~=1.17, Polymer Source Inc.) layer was $2.8\,\mathrm{nm}$ thick, as measured by ellipsometry, and could be considered as a dense polymer brush, owing to extensive previous investigations.

\textbf{Slip length measurements}

Slip lengths were measured by a velocimetry technique based on local photobleaching. The technique has been previously described in details \cite{henot_friction_2018}. We only describe here what is necessary to understand the present results. As schematically shown in Fig.~\ref{1}, a drop of polymer solution is sandwiched between two planar substrates, separated by spacers of thickness $h$. A thin line is photobleached through the fluorescent liquid while at rest. The liquid is then sheared at a constant velocity $V_\mathrm{shear}$ during a shear time $t_\mathrm{shear}$. Monitoring the evolution of the photobleached pattern under the effect of the shears allows one to directly determine the full velocity profile and to measure the slip length $b$ at the bottom surface. The real shear rate can be directly deduced from the slip length measurement:
\begin{equation}
\dot{\gamma}=\frac{V_\mathrm{shear}}{h+b}~.
\end{equation}

\section{Dependence of the slip length on the  polymer volume fraction}

The slip lengths were measured for the various  semi-dilute solutions, different shear velocities $V_\mathrm{shear}$ and different shear times $t_\mathrm{shear}$.
As recently reported \cite{grzelka_viscoelasticity-induced_2020}, a transient onset of slip at small shear times was observed for all investigated solutions. This transient onset of slip has been shown to result from the initial viscoelastic response of the solutions to the sudden onset of the shear, and to last for $5\tau_\mathrm{rep}$ \cite{grzelka_viscoelasticity-induced_2020}.
In the present paper, we focus on the steady state regime of slippage, corresponding to shear times larger than $5\tau_\mathrm{rep}$. All data presented here correspond to shear rates lower than the critical shear rate $\dot{\gamma}_c=1/\tau_\mathrm{rep}$ above which the fluid enters the shear thinning regime. The flow behavior of the semi-dilute solutions is thus only probed in their Newtonian regime.

\begin{figure}[ht]
  \centering
  \includegraphics[width=0.8\columnwidth]{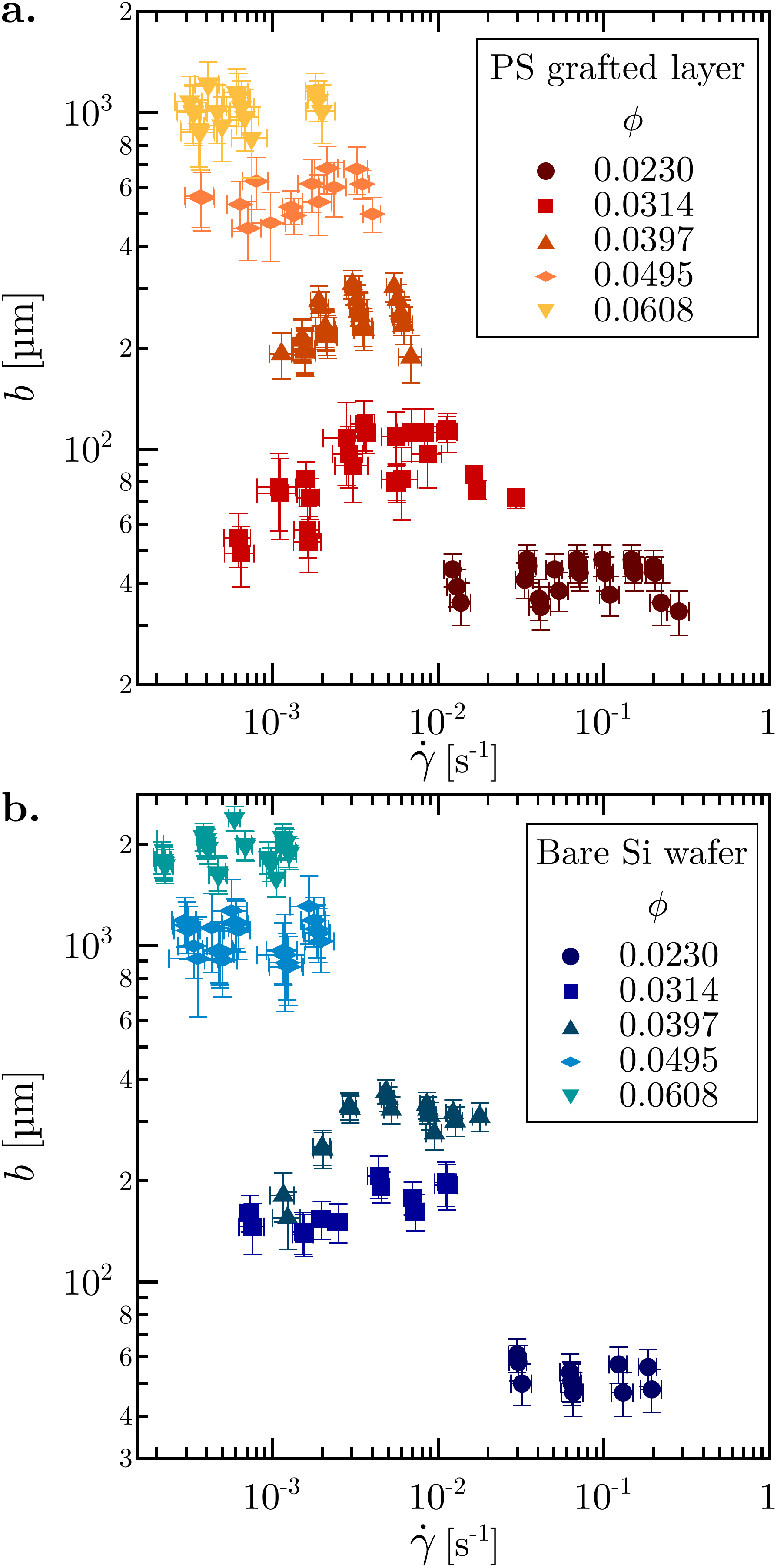}
  \caption{Slip length $b$ as a function of the real shear rate $\dot{\gamma}$ for semi-dilute solutions of high molar weight PS in DEP flowing on \textbf{(a)} grafted layer of PS brushes and \textbf{(b)} on a bare Si wafer.}
  \label{2}
\end{figure}

The data obtained for the slip length $b$ as a function of the real shear rate experienced by the solution on (a) the PS grafted layer and (b) on the bare Si wafer, are reported in Fig.~\ref{2} for shear times larger than $5\tau_\mathrm{rep}$.

Even if the polymer volume fraction $\phi$ remains small, typically only a few percent, the observed slip lengths are large, with values of the order of hundreds of micrometers and even millimeters. This underlines that slip phenomena cannot be neglected in semi-dilute polymer solutions, with orders of magnitudes for the slip lengths quite comparable to what has been reported for polymer melts \cite{migler_slip_1993, henot_friction_2018,baumchen_slip_2009}.

As can be seen in Fig.\,\ref{2}, for some of the investigated systems $\{\phi$,surface$\}$, two regimes of slip are evidenced: at low shear rates, $b$ first increases with $\dot{\gamma}$, until it reached a plateau for $\dot{\gamma}>\dot{\gamma}^*$, with the critical shear rate $\dot{\gamma}^*$ depending on both the substrate and the polymer volume fraction.
A similar transition from weak to large slip has already been observed for both melts \cite{migler_slip_1993,massey_investigation_1998,henot_sensing_2018} and polymer solutions \cite{mhetar_slip_1998, sanchez-reyes_interfacial_2003}. It has been attributed to the progressive disentanglement between the bulk flowing chains and few chains adsorbed on the substrate. For shear rates lower than $\dot{\gamma}^*$, entanglements between the surface attached chains and the bulk ones are responsible for a large interfacial friction, and thus to small slip at the wall. The surface attached chains can, however, deform under the effect of the friction forces, and become sufficiently elongated to, for large enough shear rates, be completely disentangled from bulk chains, so that high slip develops, with a high shear rate friction regime comparable to what is expected for ideal surfaces on which no adsorption takes place \cite{brochard-wyart_slippage_1996}. The presence of two slip regimes in the present experiments suggests that an adsorbed layer of PS can form, on both investigated surfaces.

We now focus on the high slip regime.
As shown in Fig.\,\ref{2}, at $\dot{\gamma}>\dot{\gamma}^*$, the slip lengths $b(\dot{\gamma})$ become essentially independent of the shear rate for each polymer volume fraction. We recall that the range of studied shear rates has been limited to the Newtonian regime for each solution, in order to avoid entering into the shear thinning regime.

\begin{figure}[htbp]
  \centering
  \includegraphics[width=\columnwidth]{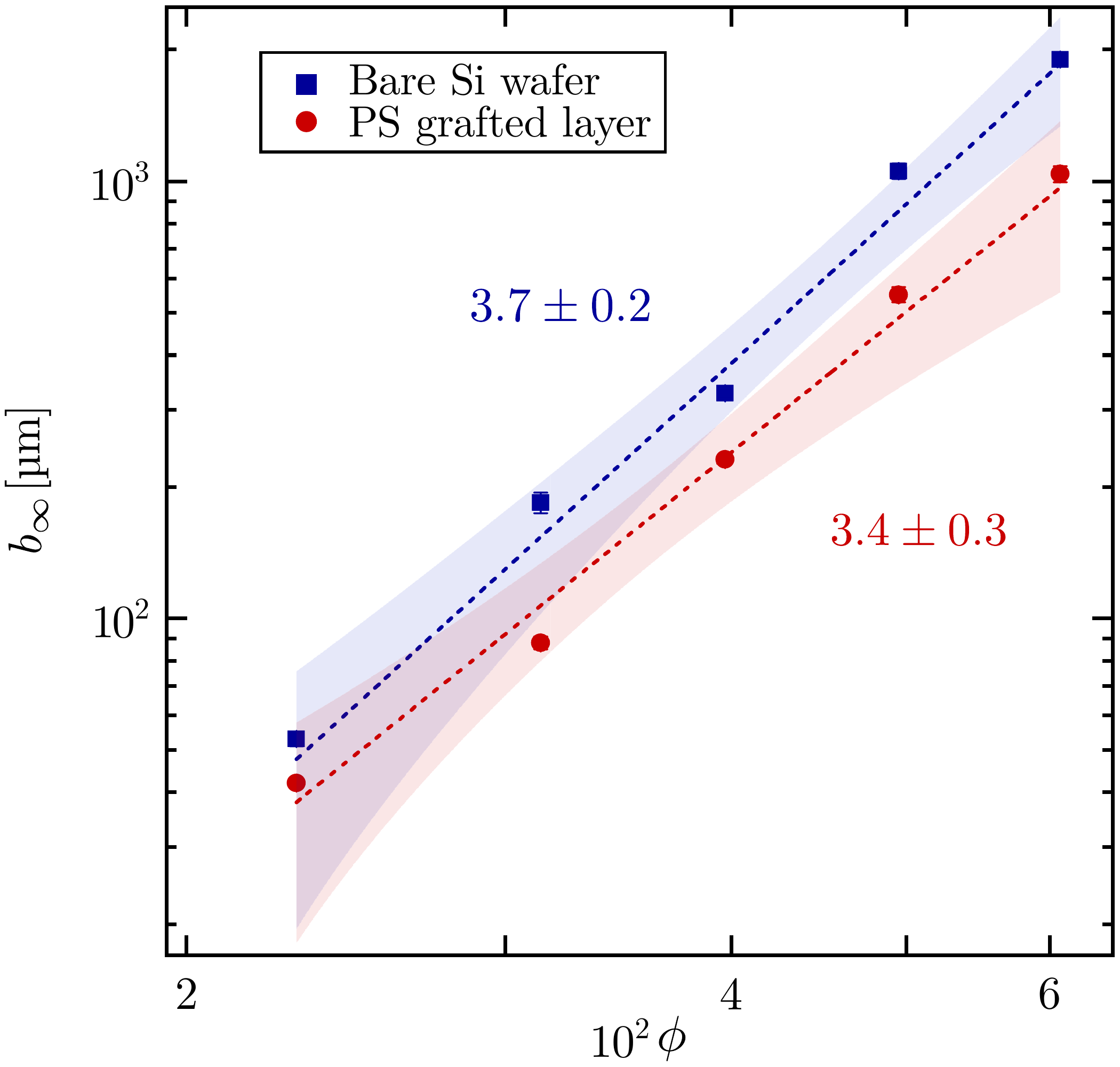}
  \caption{Mean slip length $b_\infty$ as a function of PS volume fraction $\phi$ for solutions of PS in DEP flowing on a bare Si wafer and a grafted layer of PS brushes. The dotted lines represent fits $b_\infty\propto\phi^\alpha$.}
  \label{3}
\end{figure}

As $b(\dot{\gamma})$ is constant in this high slip regime, a mean slip length $b_\infty$ at long shear times can be deduced for each polymer volume fraction.
The corresponding data are presented as a function of the polymer volume fraction $\phi$, for each substrate, in log scales, in Fig. \ref{3}.
On both substrates, $b_\infty$ increases with the polymer volume fraction $\phi$, with, at a fixed volume fraction, $b_\infty$ on the Si wafer larger than on the PS grafted layer. The dotted lines in Fig.\ref{3} are best fits to power laws, leading to $b_\mathrm{\infty,\,PS}\propto \phi^{3.4\pm0.3}$ on the PS grafted layer and to $b_\mathrm{\infty,\,Si}\propto \phi^{3.7\pm0.2}$ on the bare Si wafer. These results differ from the extrapolated data reported by Methar \textit{et al.} \cite{mhetar_slip_1998} and Sanchez-Reyes \textit{et al.}\cite{sanchez-reyes_interfacial_2003}, mainly because these last experiments were performed in the shear thinning regime of the solutions.
  The scaling exponents found on both substrates are close to each other. This suggests that the molecular mechanisms driving the interfacial friction are similar on these two model surfaces.

A first naive approach to try to rationalize these data consists in assuming that, as it has been observed in polymer melts, the interfacial friction is a local phenomenon, involving length scales comparable to the size of a monomer or of a solvent molecule (indeed, for the rather weak polymer volume fraction used, essentially solvent molecules are in contact with the substrate). It would assume $k$ to be independent of the polymer volume fraction. The slip length should then follow exactly the same scaling  as the viscosity. The scaling description of semi-dilute polymer solutions gives for the viscosity:
\begin{equation}
   \eta_\phi=\eta_\mathrm{solvent}P^3\phi^{3.75}~,
    \label{eq:scaling_viscosity}
\end{equation}with $P$ the number of monomers per polymer chain and $\eta_\mathrm{solvent}$ the solvent viscosity. This would allow predicting:
\begin{equation}
b=\frac{\eta_\mathrm{solvent}P^3\phi^{3.75}}{k}~,
\end{equation}
 The scaling exponents obtained for the volume fraction dependence of the slip length on the two surfaces are quite close to what predicts this simple naive approach but it relies on the assumption that the viscosity indeed follows the scaling prediction for polymer semi-dilute solutions, which may be questionable.

We thus have decided to independently characterize the volume fraction dependence of the viscosity (see data in supplementary material and Table\,\ref{tab6:solutions_PS}), so that the interfacial friction coefficient $k$ could be directly determined from the slip length data, relying on the Navier-de Gennes approach summarized by equation \eqref{eq:b_definition}.
Fig.\ref{4}a reports the interfacial friction coefficients thus obtained as a function of the polymer volume fraction, $k(\phi)$. One can notice that, as we focus on the Newtonian regime for all investigated solutions, the resulting friction coefficients all are independent of the shear rate in the high slip regime.

The interfacial friction coefficient is found to strongly increase with $\phi$ for both surfaces, ruling out the first naive hypothesis presented above. The dotted lines in Fig.\ref{4}a are best fits by power laws, with respectively $k_\mathrm{PS}\propto \phi^{1.9\pm0.3}$ on the PS grafted layer and $k_\mathrm{Si}\propto \phi^{1.6\pm0.2}$ on the bare Si wafer.

\section{Discussion}

\begin{figure}[htbp]
  \centering
  \includegraphics[width=0.9\columnwidth]{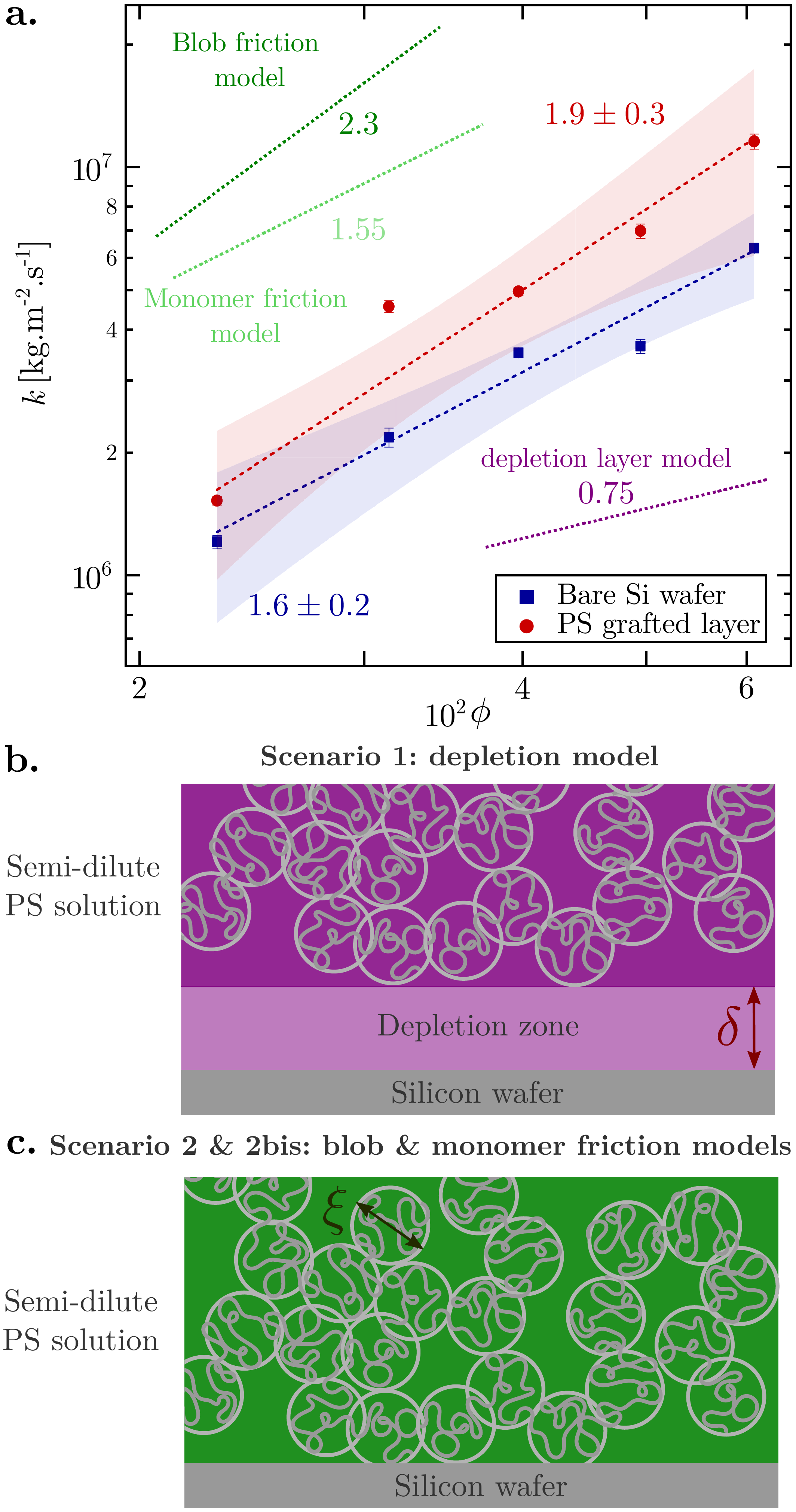}
  \caption{\textbf{(a)} Navier's coefficient $k=b_\infty/\eta$ as a function of PS volume fraction $\phi$ for solutions of PS in DEP flowing on a bare Si wafer and a grafted layer of PS brushes. The red and blue dotted lines represent fits $b_\infty\propto\phi^\beta$. The green and purple dotted lines correspond to the expected scaling laws according respectively to the blob friction model, the monomer friction model or the depletion layer model. \textbf{(b)} Illustration of the depletion model (scenario 1). $\delta$ is the thickness of the depletion layer. \textbf{(c)} Illustration of the blob friction model (scenario 2) and the monomer friction model (scenario 2bis). $\xi$ is the blob size.}
  \label{4}
\end{figure}

Two main molecular descriptions have been proposed in the past to account for flow with slip in polymer solutions, depending on the organization of the polymer chains in the immediate vicinity of the solid surface, as schematically shown in Fig.~\ref{4}b and Fig.~\ref{4}c.
We briefly present these two scenarios, and compare their predictions to our interfacial friction data.

\textbf{Scenario 1.} The first scenario applies to surfaces which are repulsive for the polymer. Then a depletion layer forms close to the interface meaning that pure solvent is in contact with the solid surface. When sheared, the polymer solution flows on top of this pure solvent layer in which a large part of the velocity gradient concentrates, due to the much lower viscosity of the solvent. No noticeable slip at the wall is expected for the solvent/solid interface, like for any simple fluid \cite{neto_boundary_2005}. The polymer solution thus only displays apparent slip. One can estimate the slip length, $b_\mathrm{dep}$, balancing the shear stresses at the solvent/polymer solution interface:
\begin{equation}
   b_\mathrm{dep}=\delta\left(\frac{\eta_\phi}{\eta_\mathrm{solvent}}-1\right)~,
     \label{eq:b_dep_eta}
\end{equation}
with $\eta_\mathrm{solvent}$ and $\eta_\phi$ the respective viscosity of the pure solvent and of the polymer solution with volume fraction $\phi$. $\delta$ is the thickness of the depletion layer. As for semi-dilute polymer solutions in good solvent $\eta_\mathrm{solvent}\ll\eta_\phi$, the slip length can be approximated as:
\begin{equation}
b_\mathrm{dep}\sim\delta\frac{\eta_\phi}{\eta_\mathrm{solvent}}~.
\label{eq:b_dep_eta2}
\end{equation}
The question is then to decide what fixes $\delta$.
For semi-dilute polymer solutions in contact with a repulsive wall, it has been predicted \cite{de_gennes_polymer_1981,joanny_effects_1979} that the thickness of the depleted layer $\delta$ should scale with the polymer volume fraction $\phi$ as the correlation length of the excluded volume interactions in the solution, $\xi$:
\begin{equation}
    \delta\sim \xi \sim a\phi^{-0.75}~,    \label{eq:scaling_dep_layer}
\end{equation}
with $a$ the Kuhn's length. Lee \textit{et al.}\cite{lee_direct_1991} and Ausserr\'e \textit{et al.}\cite{ausserre_concentration_1986} verified this scaling by direct measurements of the depleted layer for polystyrene solutions at air interface and xanthan solutions on a silica surface respectively.
This first scenario thus predicts an interfacial friction scaling as \begin{equation}
    k\sim \frac{\eta_\text{solvent}}{a}\phi^{0.75}.
    \label{eq:friction_dep_layer}
\end{equation}
\textbf{Scenario 2.} The second available scenario applies to ideal surfaces, and, relies on the scaling description of semi-dilute polymer solutions. In this framework, as long as disentanglements are not involved, the solution behaves as a closed packed suspension of blobs of size $\xi$, de-correlated from each other both for static and hydrodynamic interactions. When submitted to a shear, blobs slide on the surface. Their friction coefficient with the other blobs in the solution is the Stokes friction:
\begin{equation}
\zeta\sim {\xi \eta_\mathrm{solvent}}~.
\label{eq:frictionblob}
\end{equation}
The interfacial friction coefficient of one blob can thus be written as
\begin{equation}
  k\sim\frac{\zeta}{\xi^{2}}.
  \label{eq:friction_coed_scenario2}
\end{equation}
Both scenarios predict exactly the same scaling for the volume fraction dependence of the interfacial friction  with an exponent 0.75, which strongly differs from the measured ones. It is interesting to notice that the first scenario assumes a repulsive surface for the polymer, which is probably not the case in the present experiments, as evidenced by the observed transition from weak to large slip, indicative of the presence of an adsorbed polymer layer on the solid surface. Concerning the second scenario, it relies on the classical scaling description of the dynamics of semi-dilute polymer solutions, which validity may be questionable.

\begin{figure}[htbp]
  \centering
     \includegraphics[height=7cm]{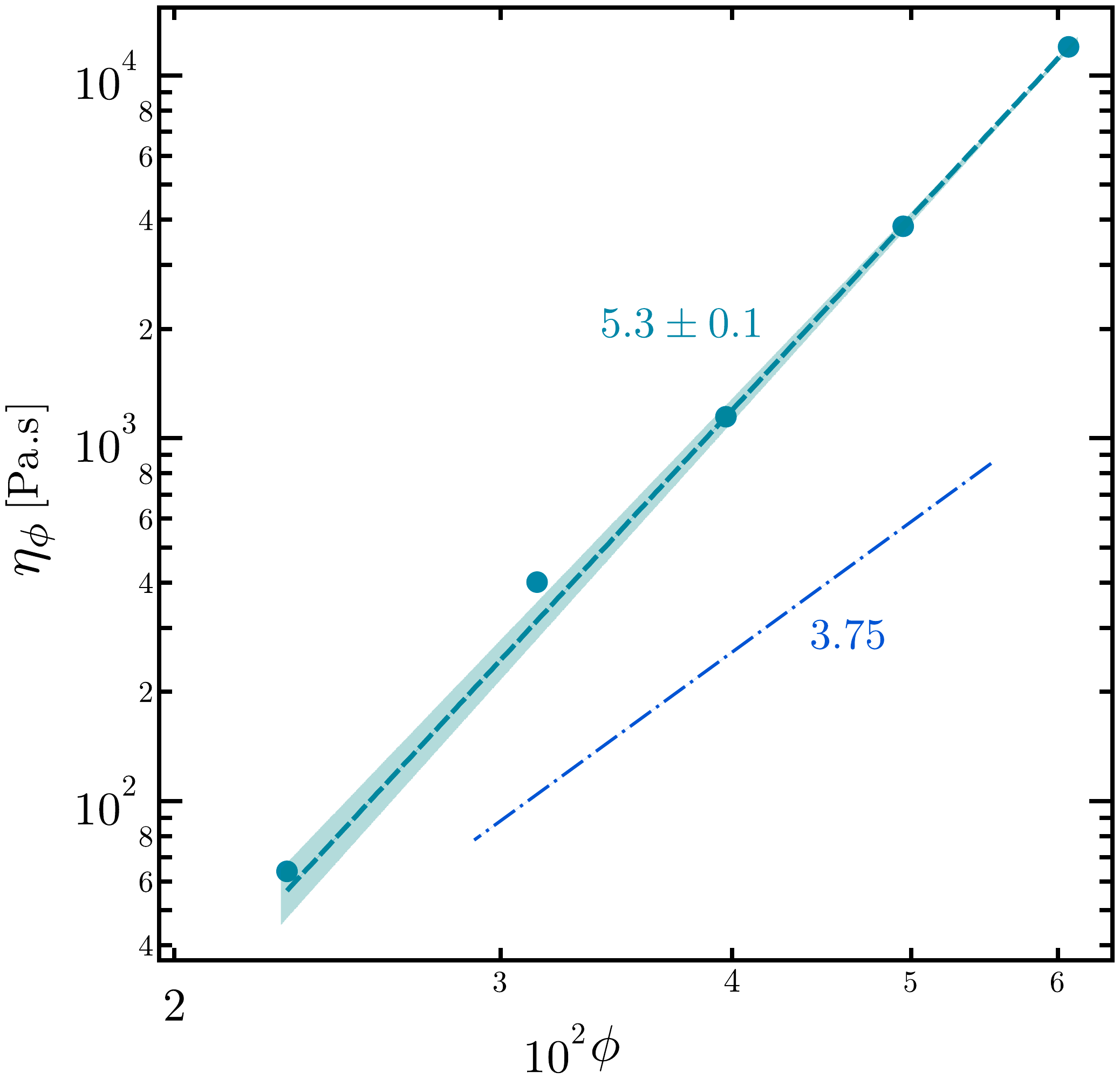}
     \caption{Zero shear viscosity $\eta_\phi$ for PS ($M_\mathrm{n} = 10.2\,\mathrm{Mg}\cdot\mathrm{mol}^{-1}$, \textit{\DJ} = 1.08) in DEP semi-diluted solutions for 5 volume fractions $\phi$. The dotted line represents the best fit, the dashed line indicates the expected scaling law.}
     \label{fig:viscosity}
\end{figure}
It is then interesting to look closer to the polymer volume fraction dependence of the viscosity that has been experimentally obtained  for the PS/DEP polymer solutions (see Fig.~\ref{fig:viscosity} and S5). The predicted scaling for the viscosity dependence versus the polymer volume fraction is clearly not obeyed (dashed line in Fig.\,\ref{fig:viscosity}), and the measured viscosities present a much steeper increase with the polymer volume fraction  (dotted line in Fig.\,\ref{fig:viscosity}) than expected from classical scaling arguments. This is not an isolated observation~\cite{osaki_relaxation_1975,mhetar_slip_1998,bhattacharjee_extensional_2002,sanchez-reyes_interfacial_2003,acharya_are_2008,henot_glissement_2018}, as visible in Fig.~S5.
Such deviations to scaling prediction have been attributed to an additional polymer volume fraction dependence of the solvent friction which appears as a prefactor in the scaling laws. Such an additional volume fraction dependence has been evidenced long ago, in particular through self diffusion measurements of small molecular probes as a function of the polymer volume fraction \cite{marmonier_reptation_1985,nemoto_concentration_1985}. When taken into consideration, these solvent friction dependencies versus the polymer volume fraction allowed one to recover good agreement with the scaling predictions. Such a sensitivity of the solvent viscosity to the polymer volume fraction has classically been attributed to the dependence of the glass transition temperature of the solutions versus the polymer volume fraction, so that, at fixed experimental temperature, the data for different polymer volume fractions correspond to  different temperature distances to the glass transition temperature of the solution $T_\mathrm{g}$ ~\cite{fox_influence_1956}. As a consequence, the solvent viscosity which appears as a prefactor in the scaling predictions for all dynamic quantities, and more specifically in equation \eqref{eq:scaling_viscosity} is not a constant quantity, but depends on the distance to $T_\mathrm{g}$ and thus on the polymer volume fraction. This applies too for the prefactor of the Stokes friction of one blob in equation \eqref{eq:frictionblob}, leading to $\zeta\sim \eta_\mathrm{solvent}(T-T_\mathrm{g}(\phi))\xi(\phi,T)$. Ferry~\cite{ferry_viscoelastic_1961} and Berry and Fox~\cite{berry_viscosity_1968} have already discussed the complex dependence of the Stokes friction on the polymer volume fraction.

Notice that we have shown through SANS experiments that the screening length of the excluded volume interactions, $\xi$ , follows the expected scaling law with $\phi$: no effect of the glass transition temperature is expected for this static quantity. Whatever the exact origin of the sensitivity of the solvent viscosity versus the polymer volume fraction, one can insert in equation \eqref{eq:frictionblob} the polymer volume fraction dependence experimentally determined from the viscosity measurements $\eta_\mathrm{solvent}(T-T_\mathrm{g}(\phi))$. This leads to an apparent power law dependence for the interfacial friction versus the polymer volume fraction with an exponent 2.3. For an easy comparison to experimental data, such a power law is reported in Fig.\,\ref{4}a as the green dotted line, and appears to be much closer to the experimental data than the 0.75 exponent predicted by the simple scaling or the depletion layer approaches.

As in the depletion layer description, pure solvent is supposed to be in contact with the solid surface, there is no reason to introduce any correction to the scaling associated to distance to the glass transition temperature varying with the polymer volume fraction. For the depletion layer, the 0.75 scaling exponent comes from the identification of the depletion layer thickness to the correlation length of the excluded volume interactions in the solution, which is a static quantity and thus insensitive to glass transition temperature effects, or any other specific dynamic effects.

It is then tempting to conclude that the observed polymer volume fraction dependence of the interfacial friction in this particular case is the signature of a friction mechanism associated to blobs sliding on the surface when subjected to shear, in the specific conditions of our experiments (large enough shear times and large enough shear rates so that the effect of the few polymer chains adsorbed at the interface can be neglected).

\textbf{Scenario 2bis.} An additional scenario can however be imagined, slightly differing from scenario 2, while starting from very similar basis, which are sketched in eq.\,\eqref{eq:friction_coed_scenario2}. This new scenario, that we call 2bis, just because it is quite close to scenario 2 in its principles, can be summarized as follows: first, as either solvent molecules or monomers are in contact with the surface, it seems not unreasonable to assume that the interfacial friction can be evaluated at the scale of a monomer or of a solvent molecule, $a$.  This is then the length scale to be entered in equations \eqref{eq:frictionblob} and \eqref{eq:friction_coed_scenario2} instead of the screening length of the hydrodynamic interactions $\xi$. This is why we call scenario 2 bis, the "monomer friction model". We have shown experimentally however that the solvent friction was depending on the polymer volume fraction. That solvent viscosity $\eta_\mathrm{solvent}(\phi)$ is the prefactor which enters into eq. \eqref{eq:frictionblob}, and its dependence in the polymer volume fraction will thus induce a dependence of the interfacial friction coefficient $k$ on the polymer volume fraction. The non local character of the interfacial friction, is,  in this scenario, assumed to only come from the solvent viscosity. The corresponding scaling, with a 1.55 scaling exponent, is shown in light dotted green in Fig.\,\ref{4}. It appears to be quite close to the friction data on the bare silicon wafer, while the scaling corresponding to the blob friction model appears closer to the friction data obtained on the grafted brush layer, so that we have no clear indications allowing one to chose between scenario 2 and 2bis. The larger friction coefficient systematically observed on the PS grafted substrate, could be attributed to the structural and dynamical features of the adsorbed polymer layer in the vicinity of the interface which is not taken into account here. One can notice however that there is a kind of intrinsic contradiction in scenario 2bis, as locality at the scale of the monomer is assumed, while non locality is introduced through the polymer volume fraction dependence of the solvent viscosity. This could possibly be justified provided the fundamental origin of such a dependence  could be fully identified. This would certainly need additional work, including systematic exploration of the temperature dependence, in order to evaluate the possible incidence of glass temperature effects, along with direct measurements of the local and semi-local friction, for example through self diffusion measurements of small molecular probes of various sizes, as a function of the over all polymer volume fraction. Part of such a program is presently underway.

Interestingly enough, this means that, for semi-dilute polymer solutions in this case, and whatever the exact origin of this effect, the interfacial friction is not only ruled by local friction phenomena developing at the scale of a monomer or of a solvent molecule. It is rather ruled by semi-local phenomena, developing at scales larger than a monomer, for which the screening length for the hydrodynamic interactions in the solution may play a role. Our data, however, point out clearly that this screening length is not the only length scale entering into play. Essentially solvent molecules are in contact with the solid surface and their whole dynamics is affected by the presence of the polymer, through two effects: the hydrodynamic correlations at the scale of the blob, and, more locally, by the fact that the whole solvent dynamics is affected by the polymer, possibly through the proximity of the glass transition temperature of the solution, which strongly depends on the polymer volume fraction.

\section{Conclusion}

Thanks to a velocimetry technique based on local fluorescence photobleaching, we have measured in a systematic manner how the polymer volume fraction was affecting interfacial slip and interfacial friction in model semi-dilute PS-DEP solutions sheared on model surfaces. Using independent measurements of the viscosity of the solutions, we  determined the Navier-de Gennes friction coefficient at the fluid-solid interface. The observed strong dependence of the interfacial friction coefficient versus the polymer volume fraction is the signature of a friction mechanism ruled by semi-local length scales. It can be interpreted in terms of blobs sliding on the surface, meaning that the important length scale is the correlation length of the hydrodynamic interactions in the polymer solution. The interfacial friction appears, however, to be a more subtle quantity: for the particular polymer solvent system used, additional polymer volume fraction effects have to be considered on top of the classical scaling approach, due to the variation of the glass transition temperature with the polymer volume fraction in the solution. Such additional effects clearly show up in the dependence of the solution viscosity versus the polymer volume fraction. If introduced as corrections to scaling, they lead to a neat improvement of the description of the experimental data for the volume fraction dependence of the interfacial friction, compared to available models. More extensive experiments will certainly be needed on other polymer/solvent systems to fully confirm the present interpretation. It seems clear, however, that, even if essentially solvent molecules are in contact with the solid surface, the polymer organization inside the solution do fix the semi-local length scale which drives interfacial friction in polymer solutions. This gives tools allowing one to manipulate this interfacial friction, and the level of slip at the wall by adjusting the polymer volume fraction.

{\bf Acknowledgments:} This  work  was  supported  by  ANR-ENCORE program (ANR-15-CE06-005). We strongly benefited from discussions with Stéphane Jouenne, Joshua McGraw, Alexandre Vilquin and François Boué. We thank Annie Brulet for her help on SANS experiments.

\bibliography{article_these_concentration_v2}

\begin{thebibliography}{49}%
\makeatletter
\providecommand \@ifxundefined [1]{%
 \@ifx{#1\undefined}
}%
\providecommand \@ifnum [1]{%
 \ifnum #1\expandafter \@firstoftwo
 \else \expandafter \@secondoftwo
 \fi
}%
\providecommand \@ifx [1]{%
 \ifx #1\expandafter \@firstoftwo
 \else \expandafter \@secondoftwo
 \fi
}%
\providecommand \natexlab [1]{#1}%
\providecommand \enquote  [1]{``#1''}%
\providecommand \bibnamefont  [1]{#1}%
\providecommand \bibfnamefont [1]{#1}%
\providecommand \citenamefont [1]{#1}%
\providecommand \href@noop [0]{\@secondoftwo}%
\providecommand \href [0]{\begingroup \@sanitize@url \@href}%
\providecommand \@href[1]{\@@startlink{#1}\@@href}%
\providecommand \@@href[1]{\endgroup#1\@@endlink}%
\providecommand \@sanitize@url [0]{\catcode `\\12\catcode `\$12\catcode
  `\&12\catcode `\#12\catcode `\^12\catcode `\_12\catcode `\%12\relax}%
\providecommand \@@startlink[1]{}%
\providecommand \@@endlink[0]{}%
\providecommand \url  [0]{\begingroup\@sanitize@url \@url }%
\providecommand \@url [1]{\endgroup\@href {#1}{\urlprefix }}%
\providecommand \urlprefix  [0]{URL }%
\providecommand \Eprint [0]{\href }%
\providecommand \doibase [0]{https://doi.org/}%
\providecommand \selectlanguage [0]{\@gobble}%
\providecommand \bibinfo  [0]{\@secondoftwo}%
\providecommand \bibfield  [0]{\@secondoftwo}%
\providecommand \translation [1]{[#1]}%
\providecommand \BibitemOpen [0]{}%
\providecommand \bibitemStop [0]{}%
\providecommand \bibitemNoStop [0]{.\EOS\space}%
\providecommand \EOS [0]{\spacefactor3000\relax}%
\providecommand \BibitemShut  [1]{\csname bibitem#1\endcsname}%
\let\auto@bib@innerbib\@empty
\bibitem [{\citenamefont {Denn}()}]{denn_extrusion_2001}%
  \BibitemOpen
  \bibfield  {author} {\bibinfo {author} {\bibfnamefont {M.}~\bibnamefont
  {Denn}},\ }\bibfield  {title} {\bibinfo {title} {Extrusion instabilities and
  wall slip},\ }\href {https://doi.org/10.1146/annurev.fluid.33.1.265} {\
  \textbf {\bibinfo {volume} {33}},\ \bibinfo {pages} {265}}\BibitemShut
  {NoStop}%
\bibitem [{\citenamefont {Stone}\ and\ \citenamefont
  {Kim}()}]{stone_microfluidics_2001}%
  \BibitemOpen
  \bibfield  {author} {\bibinfo {author} {\bibfnamefont {H.~A.}\ \bibnamefont
  {Stone}}\ and\ \bibinfo {author} {\bibfnamefont {S.}~\bibnamefont {Kim}},\
  }\bibfield  {title} {\bibinfo {title} {Microfluidics: Basic issues,
  applications, and challenges},\ }\href
  {https://doi.org/10.1002/aic.690470602} {\ \textbf {\bibinfo {volume} {47}},\
  \bibinfo {pages} {1250}}\BibitemShut {NoStop}%
\bibitem [{\citenamefont {Lauga}\ \emph {et~al.}()\citenamefont {Lauga},
  \citenamefont {Brenner},\ and\ \citenamefont
  {Stone}}]{lauga_microfluidics_2005}%
  \BibitemOpen
  \bibfield  {author} {\bibinfo {author} {\bibfnamefont {E.}~\bibnamefont
  {Lauga}}, \bibinfo {author} {\bibfnamefont {M.}~\bibnamefont {Brenner}},\
  and\ \bibinfo {author} {\bibfnamefont {H.}~\bibnamefont {Stone}},\ }\bibfield
   {title} {\bibinfo {title} {Microfluidics: The no-slip boundary condition}\
  }(\bibinfo  {publisher} {Springer, New-York})\ pp.\ \bibinfo {pages}
  {1219--1240}\BibitemShut {NoStop}%
\bibitem [{\citenamefont {Neto}\ \emph {et~al.}()\citenamefont {Neto},
  \citenamefont {Evans}, \citenamefont {Bonaccurso}, \citenamefont {Butt},\
  and\ \citenamefont {Craig}}]{neto_boundary_2005}%
  \BibitemOpen
  \bibfield  {author} {\bibinfo {author} {\bibfnamefont {C.}~\bibnamefont
  {Neto}}, \bibinfo {author} {\bibfnamefont {D.~R.}\ \bibnamefont {Evans}},
  \bibinfo {author} {\bibfnamefont {E.}~\bibnamefont {Bonaccurso}}, \bibinfo
  {author} {\bibfnamefont {H.-J.}\ \bibnamefont {Butt}},\ and\ \bibinfo
  {author} {\bibfnamefont {V.~S.~J.}\ \bibnamefont {Craig}},\ }\bibfield
  {title} {\bibinfo {title} {Boundary slip in newtonian liquids: a review of
  experimental studies},\ }\href {https://doi.org/10.1088/0034-4885/68/12/R05}
  {\ \textbf {\bibinfo {volume} {68}},\ \bibinfo {pages} {2859}}\BibitemShut
  {NoStop}%
\bibitem [{\citenamefont {Migler}\ \emph {et~al.}()\citenamefont {Migler},
  \citenamefont {Hervet},\ and\ \citenamefont {Léger}}]{migler_slip_1993}%
  \BibitemOpen
  \bibfield  {author} {\bibinfo {author} {\bibfnamefont {K.~B.}\ \bibnamefont
  {Migler}}, \bibinfo {author} {\bibfnamefont {H.}~\bibnamefont {Hervet}},\
  and\ \bibinfo {author} {\bibfnamefont {L.}~\bibnamefont {Léger}},\
  }\bibfield  {title} {\bibinfo {title} {Slip transition of a polymer melt
  under shear-stress},\ }\href@noop {} {\ \textbf {\bibinfo {volume} {70}},\
  \bibinfo {pages} {287}}\BibitemShut {NoStop}%
\bibitem [{\citenamefont {Mhetar}\ and\ \citenamefont
  {Archer}()}]{mhetar_slip_1998}%
  \BibitemOpen
  \bibfield  {author} {\bibinfo {author} {\bibfnamefont {V.}~\bibnamefont
  {Mhetar}}\ and\ \bibinfo {author} {\bibfnamefont {L.~A.}\ \bibnamefont
  {Archer}},\ }\bibfield  {title} {\bibinfo {title} {Slip in entangled polymer
  solutions},\ }\href {http://pubs.acs.org/doi/abs/10.1021/ma971339h} {\
  \textbf {\bibinfo {volume} {31}},\ \bibinfo {pages} {6639}}\BibitemShut
  {NoStop}%
\bibitem [{\citenamefont {Wang}\ and\ \citenamefont
  {Drda}()}]{wang_stick-slip_1996}%
  \BibitemOpen
  \bibfield  {author} {\bibinfo {author} {\bibfnamefont {S.-Q.}\ \bibnamefont
  {Wang}}\ and\ \bibinfo {author} {\bibfnamefont {P.~A.}\ \bibnamefont
  {Drda}},\ }\bibfield  {title} {\bibinfo {title} {Stick-slip transition in
  capillary flow of polyethylene. 2. molecular weight dependence and
  low-temperature anomaly},\ }\href {https://doi.org/10.1021/ma951512e} {\
  \textbf {\bibinfo {volume} {29}},\ \bibinfo {pages} {4115}}\BibitemShut
  {NoStop}%
\bibitem [{\citenamefont {Bäumchen}\ and\ \citenamefont
  {Jacobs}()}]{baumchen_slip_2009}%
  \BibitemOpen
  \bibfield  {author} {\bibinfo {author} {\bibfnamefont {O.}~\bibnamefont
  {Bäumchen}}\ and\ \bibinfo {author} {\bibfnamefont {K.}~\bibnamefont
  {Jacobs}},\ }\bibfield  {title} {\bibinfo {title} {Slip effects in polymer
  thin films},\ }\href {https://doi.org/10.1088/0953-8984/22/3/033102} {\
  \textbf {\bibinfo {volume} {22}},\ \bibinfo {pages} {033102}},\ \bibinfo
  {note} {publisher: {IOP} Publishing}\BibitemShut {NoStop}%
\bibitem [{\citenamefont {Granick}\ \emph {et~al.}()\citenamefont {Granick},
  \citenamefont {Zhu},\ and\ \citenamefont {Lee}}]{granick_slippery_2003}%
  \BibitemOpen
  \bibfield  {author} {\bibinfo {author} {\bibfnamefont {S.}~\bibnamefont
  {Granick}}, \bibinfo {author} {\bibfnamefont {Y.}~\bibnamefont {Zhu}},\ and\
  \bibinfo {author} {\bibfnamefont {H.}~\bibnamefont {Lee}},\ }\bibfield
  {title} {\bibinfo {title} {Slippery questions about complex fluids flowing
  past solids},\ }\href {https://doi.org/10.1038/nmat854} {\ \textbf {\bibinfo
  {volume} {2}},\ \bibinfo {pages} {221}}\BibitemShut {NoStop}%
\bibitem [{\citenamefont {Sanchez-Reyes}\ and\ \citenamefont
  {Archer}()}]{sanchez-reyes_interfacial_2003}%
  \BibitemOpen
  \bibfield  {author} {\bibinfo {author} {\bibfnamefont {J.}~\bibnamefont
  {Sanchez-Reyes}}\ and\ \bibinfo {author} {\bibfnamefont {L.~A.}\ \bibnamefont
  {Archer}},\ }\bibfield  {title} {\bibinfo {title} {Interfacial slip
  violations in polymer solutions: Role of microscale surface roughness},\
  }\href {https://doi.org/10.1021/la0265326} {\ \textbf {\bibinfo {volume}
  {19}},\ \bibinfo {pages} {3304}}\BibitemShut {NoStop}%
\bibitem [{\citenamefont {Navier}()}]{navier_memoire_1823}%
  \BibitemOpen
  \bibfield  {author} {\bibinfo {author} {\bibfnamefont {C.~L.}\ \bibnamefont
  {Navier}},\ }\bibfield  {title} {\bibinfo {title} {Mémoire sur les lois du
  mouvement des fluides},\ }in\ \href@noop {} {\emph {\bibinfo {booktitle}
  {Mémoire de l'académie des sciences de l'institut de France}}}\ (\bibinfo
  {publisher} {Imprimerie royale})\ pp.\ \bibinfo {pages}
  {389--440}\BibitemShut {NoStop}%
\bibitem [{\citenamefont
  {De~Gennes}({\natexlab{a}})}]{de_gennes_ecoulements_1979}%
  \BibitemOpen
  \bibfield  {author} {\bibinfo {author} {\bibfnamefont {P.-G.}\ \bibnamefont
  {De~Gennes}},\ }\bibfield  {title} {\bibinfo {title} {Ecoulements
  viscométriques de polymères enchevêtrés},\ }\href
  {https://doi.org/10.1142/9789812564849_0019} {\ \textbf {\bibinfo {volume}
  {288B}},\ \bibinfo {pages} {219} ({\natexlab{a}})}\BibitemShut {NoStop}%
\bibitem [{\citenamefont {Reiter}\ and\ \citenamefont
  {Khanna}()}]{reiter_real-time_2000}%
  \BibitemOpen
  \bibfield  {author} {\bibinfo {author} {\bibfnamefont {G.}~\bibnamefont
  {Reiter}}\ and\ \bibinfo {author} {\bibfnamefont {R.}~\bibnamefont
  {Khanna}},\ }\bibfield  {title} {\bibinfo {title} {Real-time determination of
  the slippage length in autophobic polymer dewetting},\ }\href
  {https://doi.org/10.1103/PhysRevLett.85.2753} {\ \textbf {\bibinfo {volume}
  {85}},\ \bibinfo {pages} {2753}}\BibitemShut {NoStop}%
\bibitem [{\citenamefont {Piau}\ and\ \citenamefont
  {El~Kissi}()}]{piau_measurement_1994}%
  \BibitemOpen
  \bibfield  {author} {\bibinfo {author} {\bibfnamefont {J.~M.}\ \bibnamefont
  {Piau}}\ and\ \bibinfo {author} {\bibfnamefont {N.}~\bibnamefont
  {El~Kissi}},\ }\bibfield  {title} {\bibinfo {title} {Measurement and
  modelling of friction in polymer melts during macroscopic slip at the wall},\
  }\href {https://doi.org/10.1016/0377-0257(94)80018-9} {\ \textbf {\bibinfo
  {volume} {54}},\ \bibinfo {pages} {121}}\BibitemShut {NoStop}%
\bibitem [{\citenamefont {El~Kissi}\ and\ \citenamefont
  {Piau}()}]{el_kissi_different_1990}%
  \BibitemOpen
  \bibfield  {author} {\bibinfo {author} {\bibfnamefont {N.}~\bibnamefont
  {El~Kissi}}\ and\ \bibinfo {author} {\bibfnamefont {J.-M.}\ \bibnamefont
  {Piau}},\ }\bibfield  {title} {\bibinfo {title} {The different capillary flow
  regimes of entangled polydimethylsiloxane polymers: macroscopic slip at the
  wall, hysteresis and cork flow},\ }\href
  {https://doi.org/10.1016/0377-0257(90)80004-J} {\ \textbf {\bibinfo {volume}
  {37}},\ \bibinfo {pages} {55}},\ \bibinfo {note} {publisher:
  Elsevier}\BibitemShut {NoStop}%
\bibitem [{\citenamefont {{McGraw}}\ \emph {et~al.}()\citenamefont {{McGraw}},
  \citenamefont {Chan}, \citenamefont {Maurer}, \citenamefont {Salez},
  \citenamefont {Benzaquen}, \citenamefont {Raphaël}, \citenamefont
  {Brinkmann},\ and\ \citenamefont {Jacobs}}]{mcgraw_slip-mediated_2016}%
  \BibitemOpen
  \bibfield  {author} {\bibinfo {author} {\bibfnamefont {J.~D.}\ \bibnamefont
  {{McGraw}}}, \bibinfo {author} {\bibfnamefont {T.~S.}\ \bibnamefont {Chan}},
  \bibinfo {author} {\bibfnamefont {S.}~\bibnamefont {Maurer}}, \bibinfo
  {author} {\bibfnamefont {T.}~\bibnamefont {Salez}}, \bibinfo {author}
  {\bibfnamefont {M.}~\bibnamefont {Benzaquen}}, \bibinfo {author}
  {\bibfnamefont {E.}~\bibnamefont {Raphaël}}, \bibinfo {author}
  {\bibfnamefont {M.}~\bibnamefont {Brinkmann}},\ and\ \bibinfo {author}
  {\bibfnamefont {K.}~\bibnamefont {Jacobs}},\ }\bibfield  {title} {\bibinfo
  {title} {Slip-mediated dewetting of polymer microdroplets},\ }\href
  {https://doi.org/10.1073/pnas.1513565113} {\ \textbf {\bibinfo {volume}
  {113}},\ \bibinfo {pages} {1168}}\BibitemShut {NoStop}%
\bibitem [{\citenamefont {Hénot}\ \emph {et~al.}({\natexlab{a}})\citenamefont
  {Hénot}, \citenamefont {Chennevière}, \citenamefont {Drockenmuller},
  \citenamefont {Léger},\ and\ \citenamefont
  {Restagno}}]{henot_comparison_2017}%
  \BibitemOpen
  \bibfield  {author} {\bibinfo {author} {\bibfnamefont {M.}~\bibnamefont
  {Hénot}}, \bibinfo {author} {\bibfnamefont {A.}~\bibnamefont
  {Chennevière}}, \bibinfo {author} {\bibfnamefont {E.}~\bibnamefont
  {Drockenmuller}}, \bibinfo {author} {\bibfnamefont {L.}~\bibnamefont
  {Léger}},\ and\ \bibinfo {author} {\bibfnamefont {F.}~\bibnamefont
  {Restagno}},\ }\bibfield  {title} {\bibinfo {title} {Comparison of the slip
  of a {PDMS} melt on weakly adsorbing surfaces measured by a new
  photobleaching-based technique},\ }\href
  {https://doi.org/10.1021/acs.macromol.7b00601} {\ \textbf {\bibinfo {volume}
  {50}},\ \bibinfo {pages} {5592} ({\natexlab{a}})}\BibitemShut {NoStop}%
\bibitem [{\citenamefont {Hénot}\ \emph {et~al.}({\natexlab{b}})\citenamefont
  {Hénot}, \citenamefont {Drockenmuller}, \citenamefont {Léger},\ and\
  \citenamefont {Restagno}}]{henot_friction_2018}%
  \BibitemOpen
  \bibfield  {author} {\bibinfo {author} {\bibfnamefont {M.}~\bibnamefont
  {Hénot}}, \bibinfo {author} {\bibfnamefont {E.}~\bibnamefont
  {Drockenmuller}}, \bibinfo {author} {\bibfnamefont {L.}~\bibnamefont
  {Léger}},\ and\ \bibinfo {author} {\bibfnamefont {F.}~\bibnamefont
  {Restagno}},\ }\bibfield  {title} {\bibinfo {title} {Friction of polymers:
  from {PDMS} melts to {PDMS} elastomers},\ }\href
  {https://doi.org/10.1021/acsmacrolett.7b00842} {\ \textbf {\bibinfo {volume}
  {7}},\ \bibinfo {pages} {112} ({\natexlab{b}})}\BibitemShut {NoStop}%
\bibitem [{\citenamefont {Durliat}\ \emph {et~al.}()\citenamefont {Durliat},
  \citenamefont {Hervet},\ and\ \citenamefont
  {Léger}}]{durliat_influence_1997}%
  \BibitemOpen
  \bibfield  {author} {\bibinfo {author} {\bibfnamefont {E.}~\bibnamefont
  {Durliat}}, \bibinfo {author} {\bibfnamefont {H.}~\bibnamefont {Hervet}},\
  and\ \bibinfo {author} {\bibfnamefont {L.}~\bibnamefont {Léger}},\
  }\bibfield  {title} {\bibinfo {title} {Influence of grafting density on wall
  slip of a polymer melt on a polymer brush},\ }\href
  {https://doi.org/10.1209/epl/i1997-00255-3} {\ \textbf {\bibinfo {volume}
  {38}},\ \bibinfo {pages} {383}}\BibitemShut {NoStop}%
\bibitem [{\citenamefont {Ilton}\ \emph {et~al.}()\citenamefont {Ilton},
  \citenamefont {Salez}, \citenamefont {Fowler}, \citenamefont {Rivetti},
  \citenamefont {Aly}, \citenamefont {Benzaquen}, \citenamefont {{McGraw}},
  \citenamefont {Raphaël}, \citenamefont {Dalnoki-Veress},\ and\ \citenamefont
  {Bäumchen}}]{ilton_adsorption-induced_2018}%
  \BibitemOpen
  \bibfield  {author} {\bibinfo {author} {\bibfnamefont {M.}~\bibnamefont
  {Ilton}}, \bibinfo {author} {\bibfnamefont {T.}~\bibnamefont {Salez}},
  \bibinfo {author} {\bibfnamefont {P.~D.}\ \bibnamefont {Fowler}}, \bibinfo
  {author} {\bibfnamefont {M.}~\bibnamefont {Rivetti}}, \bibinfo {author}
  {\bibfnamefont {M.}~\bibnamefont {Aly}}, \bibinfo {author} {\bibfnamefont
  {M.}~\bibnamefont {Benzaquen}}, \bibinfo {author} {\bibfnamefont {J.~D.}\
  \bibnamefont {{McGraw}}}, \bibinfo {author} {\bibfnamefont {E.}~\bibnamefont
  {Raphaël}}, \bibinfo {author} {\bibfnamefont {K.}~\bibnamefont
  {Dalnoki-Veress}},\ and\ \bibinfo {author} {\bibfnamefont {O.}~\bibnamefont
  {Bäumchen}},\ }\bibfield  {title} {\bibinfo {title} {Adsorption-induced slip
  inhibition for polymer melts on ideal substrates},\ }\href
  {https://doi.org/10.1038/s41467-018-03610-4} {\ \textbf {\bibinfo {volume}
  {9}},\ \bibinfo {pages} {1172}}\BibitemShut {NoStop}%
\bibitem [{\citenamefont {Müller}\ \emph {et~al.}()\citenamefont {Müller},
  \citenamefont {Pastorino},\ and\ \citenamefont
  {Servantie}}]{muller_flow_2008}%
  \BibitemOpen
  \bibfield  {author} {\bibinfo {author} {\bibfnamefont {M.}~\bibnamefont
  {Müller}}, \bibinfo {author} {\bibfnamefont {C.}~\bibnamefont {Pastorino}},\
  and\ \bibinfo {author} {\bibfnamefont {J.}~\bibnamefont {Servantie}},\
  }\bibfield  {title} {\bibinfo {title} {Flow, slippage and a hydrodynamic
  boundary condition of polymers at surfaces},\ }\href
  {https://doi.org/10.1088/0953-8984/20/49/494225} {\ \textbf {\bibinfo
  {volume} {20}},\ \bibinfo {pages} {494225}},\ \bibinfo {note} {publisher:
  {IOP} Publishing}\BibitemShut {NoStop}%
\bibitem [{\citenamefont {Montfort}()}]{montfort_polymer_2008}%
  \BibitemOpen
  \bibfield  {author} {\bibinfo {author} {\bibfnamefont {J.-P.}\ \bibnamefont
  {Montfort}},\ }\bibfield  {title} {\bibinfo {title} {Polymer chain dynamics
  and dynamic surface force apparatuses},\ }\href
  {https://doi.org/10.1021/ma7028488} {\ \textbf {\bibinfo {volume} {41}},\
  \bibinfo {pages} {5024}}\BibitemShut {NoStop}%
\bibitem [{\citenamefont {Gay}()}]{gay_grafted_1996}%
  \BibitemOpen
  \bibfield  {author} {\bibinfo {author} {\bibfnamefont {C.}~\bibnamefont
  {Gay}},\ }\bibfield  {title} {\bibinfo {title} {Grafted surface sheared by
  short polymers and the total entanglement threshold},\ }\href
  {https://doi.org/10.1051/jp2:1996186} {\ \textbf {\bibinfo {volume} {6}},\
  \bibinfo {pages} {335}},\ \bibinfo {note} {publisher: {EDP}
  Sciences}\BibitemShut {NoStop}%
\bibitem [{\citenamefont {Brochard-Wyart}\ \emph {et~al.}()\citenamefont
  {Brochard-Wyart}, \citenamefont {Gay},\ and\ \citenamefont
  {de~Gennes}}]{brochard-wyart_slippage_1996}%
  \BibitemOpen
  \bibfield  {author} {\bibinfo {author} {\bibfnamefont {F.}~\bibnamefont
  {Brochard-Wyart}}, \bibinfo {author} {\bibfnamefont {C.}~\bibnamefont
  {Gay}},\ and\ \bibinfo {author} {\bibfnamefont {P.-G.}\ \bibnamefont
  {de~Gennes}},\ }\bibfield  {title} {\bibinfo {title} {Slippage of polymer
  melts on grafted surfaces},\ }\href {https://doi.org/10.1021/ma950753j} {\
  \textbf {\bibinfo {volume} {29}},\ \bibinfo {pages} {377}},\ \bibinfo {note}
  {publisher: American Chemical Society}\BibitemShut {NoStop}%
\bibitem [{\citenamefont {Deng}\ \emph {et~al.}()\citenamefont {Deng},
  \citenamefont {Li}, \citenamefont {Liang}, \citenamefont {Caswell},\ and\
  \citenamefont {Karniadakis}}]{deng_simulation_2012}%
  \BibitemOpen
  \bibfield  {author} {\bibinfo {author} {\bibfnamefont {M.}~\bibnamefont
  {Deng}}, \bibinfo {author} {\bibfnamefont {X.}~\bibnamefont {Li}}, \bibinfo
  {author} {\bibfnamefont {H.}~\bibnamefont {Liang}}, \bibinfo {author}
  {\bibfnamefont {B.}~\bibnamefont {Caswell}},\ and\ \bibinfo {author}
  {\bibfnamefont {G.~E.}\ \bibnamefont {Karniadakis}},\ }\bibfield  {title}
  {\bibinfo {title} {Simulation and modelling of slip flow over surfaces
  grafted with polymer brushes and glycocalyx fibres},\ }\href
  {https://doi.org/10.1017/jfm.2012.387} {\ \textbf {\bibinfo {volume} {711}},\
  \bibinfo {pages} {192}}\BibitemShut {NoStop}%
\bibitem [{\citenamefont {Barnes}()}]{barnes_review_1995}%
  \BibitemOpen
  \bibfield  {author} {\bibinfo {author} {\bibfnamefont {H.~A.}\ \bibnamefont
  {Barnes}},\ }\bibfield  {title} {\bibinfo {title} {A review of the slip (wall
  depletion) of polymer solutions, emulsions and particle suspensions in
  viscometers: its cause, character, and cure},\ }\href
  {https://doi.org/10.1016/0377-0257(94)01282-M} {\ \textbf {\bibinfo {volume}
  {56}},\ \bibinfo {pages} {221}}\BibitemShut {NoStop}%
\bibitem [{\citenamefont {Graham}()}]{graham_fluid_2011}%
  \BibitemOpen
  \bibfield  {author} {\bibinfo {author} {\bibfnamefont {M.~D.}\ \bibnamefont
  {Graham}},\ }\bibfield  {title} {\bibinfo {title} {Fluid dynamics of
  dissolved polymer molecules in confined geometries},\ }\href
  {https://doi.org/10.1146/annurev-fluid-121108-145523} {\ \textbf {\bibinfo
  {volume} {43}},\ \bibinfo {pages} {273}}\BibitemShut {NoStop}%
\bibitem [{\citenamefont {Ma}\ and\ \citenamefont {Graham}()}]{ma_theory_2005}%
  \BibitemOpen
  \bibfield  {author} {\bibinfo {author} {\bibfnamefont {H.}~\bibnamefont
  {Ma}}\ and\ \bibinfo {author} {\bibfnamefont {M.~D.}\ \bibnamefont
  {Graham}},\ }\bibfield  {title} {\bibinfo {title} {Theory of shear-induced
  migration in dilute polymer solutions near solid boundaries},\ }\href
  {https://doi.org/10.1063/1.2011367} {\ \textbf {\bibinfo {volume} {17}},\
  \bibinfo {pages} {083103}}\BibitemShut {NoStop}%
\bibitem [{\citenamefont {Boukany}\ \emph {et~al.}()\citenamefont {Boukany},
  \citenamefont {Hemminger}, \citenamefont {Wang},\ and\ \citenamefont
  {Lee}}]{boukany_molecular_2010}%
  \BibitemOpen
  \bibfield  {author} {\bibinfo {author} {\bibfnamefont {P.~E.}\ \bibnamefont
  {Boukany}}, \bibinfo {author} {\bibfnamefont {O.}~\bibnamefont {Hemminger}},
  \bibinfo {author} {\bibfnamefont {S.-Q.}\ \bibnamefont {Wang}},\ and\
  \bibinfo {author} {\bibfnamefont {L.~J.}\ \bibnamefont {Lee}},\ }\bibfield
  {title} {\bibinfo {title} {Molecular imaging of slip in entangled {DNA}
  solution}\ }\textbf {\bibinfo {volume} {105}},\ \href
  {https://doi.org/10.1103/PhysRevLett.105.027802}
  {10.1103/PhysRevLett.105.027802}\BibitemShut {NoStop}%
\bibitem [{\citenamefont {Archer}\ \emph {et~al.}()\citenamefont {Archer},
  \citenamefont {Chen},\ and\ \citenamefont {Larson}}]{archer_delayed_1995}%
  \BibitemOpen
  \bibfield  {author} {\bibinfo {author} {\bibfnamefont {L.~A.}\ \bibnamefont
  {Archer}}, \bibinfo {author} {\bibfnamefont {Y.}~\bibnamefont {Chen}},\ and\
  \bibinfo {author} {\bibfnamefont {R.~G.}\ \bibnamefont {Larson}},\ }\bibfield
   {title} {\bibinfo {title} {Delayed slip after step strains in highly
  entangled polystyrene solutions},\ }\href {https://doi.org/10.1122/1.550710}
  {\ \textbf {\bibinfo {volume} {39}},\ \bibinfo {pages} {519}}\BibitemShut
  {NoStop}%
\bibitem [{\citenamefont {Grzelka}\ \emph {et~al.}()\citenamefont {Grzelka},
  \citenamefont {Antoniuk}, \citenamefont {Drockenmuller}, \citenamefont
  {Chennevière}, \citenamefont {Léger},\ and\ \citenamefont
  {Restagno}}]{grzelka_viscoelasticity-induced_2020}%
  \BibitemOpen
  \bibfield  {author} {\bibinfo {author} {\bibfnamefont {M.}~\bibnamefont
  {Grzelka}}, \bibinfo {author} {\bibfnamefont {I.}~\bibnamefont {Antoniuk}},
  \bibinfo {author} {\bibfnamefont {E.}~\bibnamefont {Drockenmuller}}, \bibinfo
  {author} {\bibfnamefont {A.}~\bibnamefont {Chennevière}}, \bibinfo {author}
  {\bibfnamefont {L.}~\bibnamefont {Léger}},\ and\ \bibinfo {author}
  {\bibfnamefont {F.}~\bibnamefont {Restagno}},\ }\bibfield  {title} {\bibinfo
  {title} {Viscoelasticity-induced onset of slip at the wall for polymer
  fluids},\ }\href {https://doi.org/10.1021/acsmacrolett.0c00182} {\ ,\
  \bibinfo {pages} {924}}\BibitemShut {NoStop}%
\bibitem [{\citenamefont {Barraud}\ \emph {et~al.}({\natexlab{a}})\citenamefont
  {Barraud}, \citenamefont {Cross}, \citenamefont {Picard}, \citenamefont
  {Restagno}, \citenamefont {Léger},\ and\ \citenamefont
  {Charlaix}}]{barraud_boundary_2018}%
  \BibitemOpen
  \bibfield  {author} {\bibinfo {author} {\bibfnamefont {C.}~\bibnamefont
  {Barraud}}, \bibinfo {author} {\bibfnamefont {B.}~\bibnamefont {Cross}},
  \bibinfo {author} {\bibfnamefont {C.}~\bibnamefont {Picard}}, \bibinfo
  {author} {\bibfnamefont {F.}~\bibnamefont {Restagno}}, \bibinfo {author}
  {\bibfnamefont {L.}~\bibnamefont {Léger}},\ and\ \bibinfo {author}
  {\bibfnamefont {E.}~\bibnamefont {Charlaix}},\ }\bibfield  {title} {\bibinfo
  {title} {Boundary flow of viscoelastic polyelectrolyte solutions},\ }\href
  {http://arxiv.org/abs/1803.03440} {\  ({\natexlab{a}})}\BibitemShut {NoStop}%
\bibitem [{\citenamefont {Barraud}\ \emph {et~al.}({\natexlab{b}})\citenamefont
  {Barraud}, \citenamefont {Cross}, \citenamefont {Picard}, \citenamefont
  {Restagno}, \citenamefont {Léger},\ and\ \citenamefont
  {Charlaix}}]{barraud_large_2019}%
  \BibitemOpen
  \bibfield  {author} {\bibinfo {author} {\bibfnamefont {C.}~\bibnamefont
  {Barraud}}, \bibinfo {author} {\bibfnamefont {B.}~\bibnamefont {Cross}},
  \bibinfo {author} {\bibfnamefont {C.}~\bibnamefont {Picard}}, \bibinfo
  {author} {\bibfnamefont {F.}~\bibnamefont {Restagno}}, \bibinfo {author}
  {\bibfnamefont {L.}~\bibnamefont {Léger}},\ and\ \bibinfo {author}
  {\bibfnamefont {E.}~\bibnamefont {Charlaix}},\ }\bibfield  {title} {\bibinfo
  {title} {Large slippage and depletion layer at the polyelectrolyte/solid
  interface},\ }\href {https://doi.org/10.1039/C9SM00910H} {\ \textbf {\bibinfo
  {volume} {15}},\ \bibinfo {pages} {6308} ({\natexlab{b}})}\BibitemShut
  {NoStop}%
\bibitem [{\citenamefont {Chennevière}\ \emph {et~al.}()\citenamefont
  {Chennevière}, \citenamefont {Cousin}, \citenamefont {Boué}, \citenamefont
  {Drockenmuller}, \citenamefont {Shull}, \citenamefont {Léger},\ and\
  \citenamefont {Restagno}}]{chenneviere_direct_2016}%
  \BibitemOpen
  \bibfield  {author} {\bibinfo {author} {\bibfnamefont {A.}~\bibnamefont
  {Chennevière}}, \bibinfo {author} {\bibfnamefont {F.}~\bibnamefont
  {Cousin}}, \bibinfo {author} {\bibfnamefont {F.}~\bibnamefont {Boué}},
  \bibinfo {author} {\bibfnamefont {E.}~\bibnamefont {Drockenmuller}}, \bibinfo
  {author} {\bibfnamefont {K.~R.}\ \bibnamefont {Shull}}, \bibinfo {author}
  {\bibfnamefont {L.}~\bibnamefont {Léger}},\ and\ \bibinfo {author}
  {\bibfnamefont {F.}~\bibnamefont {Restagno}},\ }\bibfield  {title} {\bibinfo
  {title} {Direct molecular evidence of the origin of slip of polymer melts on
  grafted brushes},\ }\href {https://doi.org/10.1021/acs.macromol.5b02505} {\
  \textbf {\bibinfo {volume} {49}},\ \bibinfo {pages} {2348}}\BibitemShut
  {NoStop}%
\bibitem [{\citenamefont {Massey}\ \emph {et~al.}()\citenamefont {Massey},
  \citenamefont {Hervet},\ and\ \citenamefont
  {Léger}}]{massey_investigation_1998}%
  \BibitemOpen
  \bibfield  {author} {\bibinfo {author} {\bibfnamefont {G.}~\bibnamefont
  {Massey}}, \bibinfo {author} {\bibfnamefont {H.}~\bibnamefont {Hervet}},\
  and\ \bibinfo {author} {\bibfnamefont {L.}~\bibnamefont {Léger}},\
  }\bibfield  {title} {\bibinfo {title} {Investigation of the slip transition
  at the melt polymer interface},\ }\href
  {https://doi.org/10.1209/epl/i1998-00323-8} {\ \textbf {\bibinfo {volume}
  {43}},\ \bibinfo {pages} {83}}\BibitemShut {NoStop}%
\bibitem [{\citenamefont {Hénot}\ \emph {et~al.}({\natexlab{c}})\citenamefont
  {Hénot}, \citenamefont {Drockenmuller}, \citenamefont {Léger},\ and\
  \citenamefont {Restagno}}]{henot_sensing_2018}%
  \BibitemOpen
  \bibfield  {author} {\bibinfo {author} {\bibfnamefont {M.}~\bibnamefont
  {Hénot}}, \bibinfo {author} {\bibfnamefont {E.}~\bibnamefont
  {Drockenmuller}}, \bibinfo {author} {\bibfnamefont {L.}~\bibnamefont
  {Léger}},\ and\ \bibinfo {author} {\bibfnamefont {F.}~\bibnamefont
  {Restagno}},\ }\bibfield  {title} {\bibinfo {title} {Sensing adsorption
  kinetics through slip velocity measurements of polymer melts}\ }\textbf
  {\bibinfo {volume} {41}},\ \href {https://doi.org/10.1140/epje/i2018-11697-4}
  {10.1140/epje/i2018-11697-4} ({\natexlab{c}})\BibitemShut {NoStop}%
\bibitem [{\citenamefont {De~Gennes}({\natexlab{b}})}]{de_gennes_polymer_1981}%
  \BibitemOpen
  \bibfield  {author} {\bibinfo {author} {\bibfnamefont {P.-G.}\ \bibnamefont
  {De~Gennes}},\ }\bibfield  {title} {\bibinfo {title} {Polymer solutions near
  an interface. adsorption and depletion layers},\ }\href
  {https://doi.org/10.1021/ma50007a007} {\ \textbf {\bibinfo {volume} {14}},\
  \bibinfo {pages} {1637} ({\natexlab{b}})}\BibitemShut {NoStop}%
\bibitem [{\citenamefont {Joanny}\ \emph {et~al.}()\citenamefont {Joanny},
  \citenamefont {Leibler},\ and\ \citenamefont {Gennes}}]{joanny_effects_1979}%
  \BibitemOpen
  \bibfield  {author} {\bibinfo {author} {\bibfnamefont {J.~F.}\ \bibnamefont
  {Joanny}}, \bibinfo {author} {\bibfnamefont {L.}~\bibnamefont {Leibler}},\
  and\ \bibinfo {author} {\bibfnamefont {P.~G.~D.}\ \bibnamefont {Gennes}},\
  }\bibfield  {title} {\bibinfo {title} {Effects of polymer solutions on
  colloid stability},\ }\href {https://doi.org/10.1002/pol.1979.180170615} {\
  \textbf {\bibinfo {volume} {17}},\ \bibinfo {pages} {1073}}\BibitemShut
  {NoStop}%
\bibitem [{\citenamefont {Lee}\ \emph {et~al.}()\citenamefont {Lee},
  \citenamefont {Guiselin}, \citenamefont {Lapp}, \citenamefont {Farnoux},\
  and\ \citenamefont {Penfold}}]{lee_direct_1991}%
  \BibitemOpen
  \bibfield  {author} {\bibinfo {author} {\bibfnamefont {L.-T.}\ \bibnamefont
  {Lee}}, \bibinfo {author} {\bibfnamefont {O.}~\bibnamefont {Guiselin}},
  \bibinfo {author} {\bibfnamefont {A.}~\bibnamefont {Lapp}}, \bibinfo {author}
  {\bibfnamefont {B.}~\bibnamefont {Farnoux}},\ and\ \bibinfo {author}
  {\bibfnamefont {J.}~\bibnamefont {Penfold}},\ }\bibfield  {title} {\bibinfo
  {title} {Direct measurements of polymer depletion layers by neutron
  reflectivity},\ }\href {https://doi.org/10.1103/PhysRevLett.67.2838} {\
  \textbf {\bibinfo {volume} {67}},\ \bibinfo {pages} {2838}}\BibitemShut
  {NoStop}%
\bibitem [{\citenamefont {Ausserre}\ \emph {et~al.}()\citenamefont {Ausserre},
  \citenamefont {Hervet},\ and\ \citenamefont
  {Rondelez}}]{ausserre_concentration_1986}%
  \BibitemOpen
  \bibfield  {author} {\bibinfo {author} {\bibfnamefont {D.}~\bibnamefont
  {Ausserre}}, \bibinfo {author} {\bibfnamefont {H.}~\bibnamefont {Hervet}},\
  and\ \bibinfo {author} {\bibfnamefont {F.}~\bibnamefont {Rondelez}},\
  }\bibfield  {title} {\bibinfo {title} {Concentration dependence of the
  interfacial depletion layer thickness for polymer solutions in contact with
  nonadsorbing walls},\ }\href {https://doi.org/10.1021/ma00155a015} {\ \textbf
  {\bibinfo {volume} {19}},\ \bibinfo {pages} {85}},\ \bibinfo {note}
  {publisher: American Chemical Society}\BibitemShut {NoStop}%
\bibitem [{\citenamefont {Osaki}\ \emph {et~al.}()\citenamefont {Osaki},
  \citenamefont {Fukuda},\ and\ \citenamefont
  {Kurata}}]{osaki_relaxation_1975}%
  \BibitemOpen
  \bibfield  {author} {\bibinfo {author} {\bibfnamefont {K.}~\bibnamefont
  {Osaki}}, \bibinfo {author} {\bibfnamefont {M.}~\bibnamefont {Fukuda}},\ and\
  \bibinfo {author} {\bibfnamefont {M.}~\bibnamefont {Kurata}},\ }\bibfield
  {title} {\bibinfo {title} {Relaxation spectra of concentrated polystyrene
  solutions},\ }\href {https://doi.org/10.1002/pol.1975.180130410} {\ \textbf
  {\bibinfo {volume} {13}},\ \bibinfo {pages} {775}}\BibitemShut {NoStop}%
\bibitem [{\citenamefont {Bhattacharjee}\ \emph {et~al.}()\citenamefont
  {Bhattacharjee}, \citenamefont {Oberhauser}, \citenamefont {{McKinley}},
  \citenamefont {Leal},\ and\ \citenamefont
  {Sridhar}}]{bhattacharjee_extensional_2002}%
  \BibitemOpen
  \bibfield  {author} {\bibinfo {author} {\bibfnamefont {P.~K.}\ \bibnamefont
  {Bhattacharjee}}, \bibinfo {author} {\bibfnamefont {J.~P.}\ \bibnamefont
  {Oberhauser}}, \bibinfo {author} {\bibfnamefont {G.~H.}\ \bibnamefont
  {{McKinley}}}, \bibinfo {author} {\bibfnamefont {L.~G.}\ \bibnamefont
  {Leal}},\ and\ \bibinfo {author} {\bibfnamefont {T.}~\bibnamefont
  {Sridhar}},\ }\bibfield  {title} {\bibinfo {title} {Extensional rheometry of
  entangled solutions},\ }\href {https://doi.org/10.1021/ma0118623} {\ \textbf
  {\bibinfo {volume} {35}},\ \bibinfo {pages} {10131}}\BibitemShut {NoStop}%
\bibitem [{\citenamefont {Acharya}\ \emph {et~al.}()\citenamefont {Acharya},
  \citenamefont {Bhattacharjee}, \citenamefont {Nguyen},\ and\ \citenamefont
  {Sridhar}}]{acharya_are_2008}%
  \BibitemOpen
  \bibfield  {author} {\bibinfo {author} {\bibfnamefont {M.~V.}\ \bibnamefont
  {Acharya}}, \bibinfo {author} {\bibfnamefont {P.~K.}\ \bibnamefont
  {Bhattacharjee}}, \bibinfo {author} {\bibfnamefont {D.~A.}\ \bibnamefont
  {Nguyen}},\ and\ \bibinfo {author} {\bibfnamefont {T.}~\bibnamefont
  {Sridhar}},\ }\bibfield  {title} {\bibinfo {title} {Are entangled polymeric
  solutions different from melts?},\ }\href {https://doi.org/10.1063/1.2964702}
  {\ \textbf {\bibinfo {volume} {1027}},\ \bibinfo {pages} {391}}\BibitemShut
  {NoStop}%
\bibitem [{\citenamefont {Hénot}()}]{henot_glissement_2018}%
  \BibitemOpen
  \bibfield  {author} {\bibinfo {author} {\bibfnamefont {M.}~\bibnamefont
  {Hénot}},\ }\emph {\bibinfo {title} {Glissement de polymères liquides}},\
  \href {http://www.theses.fr/2018SACLS194} {\bibinfo {type} {Phd dissertation
  thesis}},\ \bibinfo {note} {paris Saclay}\BibitemShut {NoStop}%
\bibitem [{\citenamefont {Marmonier}\ and\ \citenamefont
  {Léger}(1985)}]{marmonier_reptation_1985}%
  \BibitemOpen
  \bibfield  {author} {\bibinfo {author} {\bibfnamefont {M.~F.}\ \bibnamefont
  {Marmonier}}\ and\ \bibinfo {author} {\bibfnamefont {L.}~\bibnamefont
  {Léger}},\ }\bibfield  {title} {\bibinfo {title} {Reptation and {Tube}
  {Renewal} in {Entangled} {Polymer} {Solutions}},\ }\href
  {https://doi.org/10.1103/PhysRevLett.55.1078} {\bibfield  {journal} {\bibinfo
   {journal} {Physical Review Letters}\ }\textbf {\bibinfo {volume} {55}},\
  \bibinfo {pages} {1078} (\bibinfo {year} {1985})},\ \bibinfo {note}
  {publisher: American Physical Society}\BibitemShut {NoStop}%
\bibitem [{\citenamefont {Nemoto}\ \emph {et~al.}(1985)\citenamefont {Nemoto},
  \citenamefont {Landry}, \citenamefont {Noh}, \citenamefont {Kitano},
  \citenamefont {Wesson},\ and\ \citenamefont
  {Yu}}]{nemoto_concentration_1985}%
  \BibitemOpen
  \bibfield  {author} {\bibinfo {author} {\bibfnamefont {N.}~\bibnamefont
  {Nemoto}}, \bibinfo {author} {\bibfnamefont {M.~R.}\ \bibnamefont {Landry}},
  \bibinfo {author} {\bibfnamefont {I.}~\bibnamefont {Noh}}, \bibinfo {author}
  {\bibfnamefont {T.}~\bibnamefont {Kitano}}, \bibinfo {author} {\bibfnamefont
  {J.~A.}\ \bibnamefont {Wesson}},\ and\ \bibinfo {author} {\bibfnamefont
  {H.}~\bibnamefont {Yu}},\ }\bibfield  {title} {\bibinfo {title}
  {Concentration dependence of self-diffusion coefficient by forced {Rayleigh}
  scattering: polystyrene in tetrahydrofuran},\ }\href
  {https://doi.org/10.1021/ma00144a038} {\bibfield  {journal} {\bibinfo
  {journal} {Macromolecules}\ }\textbf {\bibinfo {volume} {18}},\ \bibinfo
  {pages} {308} (\bibinfo {year} {1985})},\ \bibinfo {note} {publisher:
  American Chemical Society}\BibitemShut {NoStop}%
\bibitem [{\citenamefont {Fox}()}]{fox_influence_1956}%
  \BibitemOpen
  \bibfield  {author} {\bibinfo {author} {\bibfnamefont {T.~G.}\ \bibnamefont
  {Fox}},\ }\bibfield  {title} {\bibinfo {title} {Influence of diluent and of
  copolymer composition on the glass temperature of a polymer system},\ }\href
  {https://ci.nii.ac.jp/naid/10006149119/} {\ \textbf {\bibinfo {volume} {1}},\
  \bibinfo {pages} {123}}\BibitemShut {NoStop}%
\bibitem [{\citenamefont {Ferry}()}]{ferry_viscoelastic_1961}%
  \BibitemOpen
  \bibfield  {author} {\bibinfo {author} {\bibfnamefont {J.~D.}\ \bibnamefont
  {Ferry}},\ }\href@noop {} {\emph {\bibinfo {title} {Viscoelastic Properties
  of Polymers}}}\ (\bibinfo  {publisher} {Wiley})\BibitemShut {NoStop}%
\bibitem [{\citenamefont {Berry}\ and\ \citenamefont
  {Fox}()}]{berry_viscosity_1968}%
  \BibitemOpen
  \bibfield  {author} {\bibinfo {author} {\bibfnamefont {G.~C.}\ \bibnamefont
  {Berry}}\ and\ \bibinfo {author} {\bibfnamefont {T.~G.}\ \bibnamefont
  {Fox}},\ }\bibfield  {title} {\bibinfo {title} {The viscosity of polymers and
  their concentrated solutions},\ }in\ \href
  {https://link.springer.com/chapter/10.1007/BFb0050985} {\emph {\bibinfo
  {booktitle} {Fortschritte der Hochpolymeren-Forschung}}},\ \bibinfo {series}
  {Advances in Polymer Science}, Vol.~\bibinfo {volume} {5}\ (\bibinfo
  {publisher} {Spinger})\ pp.\ \bibinfo {pages} {261--357}\BibitemShut
  {NoStop}%
\end{thebibliography}%
\end{document}